\documentclass[runningheads]{llncs}

\usepackage{booktabs}

\usepackage{graphicx}
\usepackage[outdir=./]{epstopdf}
\usepackage{tikz}

\usepackage[linesnumbered,ruled,noend]{algorithm2e}
\usepackage{listings}
\usepackage{multicol}

\newlength{\commentWidth}
\newcommand{\atcp}[1]{\tcp*[r]{\makebox[\commentWidth]{\small{#1}\hfill}}}
\newcommand{\btcp}[1]{\tcp*[f]{\makebox[\commentWidth]{\small{#1}\hfill}}}

\usepackage{xcolor}
\usepackage{hyperref}

\usepackage[utf8]{inputenc}
\usepackage[english]{babel}
\usepackage{blindtext}
\usepackage{amsfonts}
\usepackage{fontawesome}
\usepackage{amsmath}
\usepackage{amssymb}

\def\etal/{et~al.}
\def\ie/{\textit{i.e.},}
\def\eg/{e.g.,}
\def\cf/{cf.}
\def\1/{\top}
\def\0/{\bot}
\def\bools/{\mathbb{B}}
\def\tool/{\textit{Danira}}
\def\keccak/{\textsc{Keccak}}
\def\rebecca/{\textsc{Rebecca}}
\def\coco/{\textsc{Coco}}

\newcommand{\sub}[2]{\left[ #2 \leftarrow #1 \right]}
\newcommand{\app}[1]{\left[ #1 \right]}
\newcommand{\para}[1]{\textbf{\textit{#1}}\ }
\newcommand{\ess}[1]{\mathcal{D}(#1)}
\newcommand{\xess}[1]{\widehat{\mathcal{D}}(#1)}
\newcommand{\numass}[2]{\text{\small\#}_{#1}(#2)}
\newcommand{\fvar}[1]{\operatorname{Var}(#1)}
\newcommand{\pow}[1]{\mathcal{P}(#1)}
\newcommand{\fnl}[1]{{#1}^{\operatorname{nl}}}
\newcommand{\flin}[1]{{#1}^{\operatorname{lin}}}
\newcommand{\fact}[1]{\mathcal{C}(#1)}
\newcommand{\xfact}[1]{\widehat{\mathcal{C}}(#1)}

\def\secrets/{S}
\def\masks/{M}

\ifdefined\print
\usepackage[appendix=strip]{apxproof}
\else
\usepackage[appendix=inline]{apxproof}
\fi

\newtheoremrep{lemma}{Lemma}
\newtheoremrep{theorem}{Theorem}

\title{Proving SIFA Protection of Masked Redundant Circuits
\thanks{This work was supported by the \emph{Austrian Research Promotion Agency
(FFG)} via the \textbf{FERMION} project (grant number 867542, ICT of the Future), and the K-project \textbf{DeSSnet} (funded in the context of COMET).
\ifdefined\print
We refer to the online appendix~\cite{paper-appendix} for the formal proofs of the presented lemmas and theorems.
\else
The final authenticated version is available online at \url{https://doi.org/[Insert DOI]}.
\fi
}
}

\author{Vedad Hadžić \and 
Robert Primas \and
Roderick Bloem}

\institute{Graz University of Technology, Graz, Austria\\
\email{firstname.lastname@iaik.tugraz.at}\\
\url{https://www.iaik.tugraz.at/}}

\begin{document}

\maketitle

\vspace{-0.5cm}
\begin{abstract}
Implementation attacks like side-channel and fault attacks pose a considerable threat
to cryptographic devices that are physically accessible by an attacker.
As a consequence, devices like smart cards implement corresponding
countermeasures like redundant computation and masking.
Recently, statistically ineffective fault attacks (SIFA) were shown to be able to circumvent
these classical countermeasure techniques.
We present a new approach for verifying the SIFA protection of arbitrary masked implementations in both hardware and software.
The proposed method uses Boolean dependency analysis, factorization, and known properties of
masked computations to show whether the fault detection mechanism of redundant masked circuits can leak information about the processed secret values.
We implemented this new method in a tool called \tool/, which can show the SIFA resistance of cryptographic implementations like AES S-Boxes within minutes. 

\end{abstract}

\section{Introduction}

Cryptographic primitives are primarily designed to withstand mathematical attacks in a black-box setting. 
However, when these primitives are deployed in the real world, they find themselves in a grey-box setting in which an attacker may try to force faulty computations or observe additional physical side-channel information, such as instantaneous power consumption. 
This improved attacker capability simplifies the extraction of secrets like cryptographic keys. 

Active implementation attacks, such as fault analysis~\cite{eurocryptBonehDL97,cryptoBihamS97}, and passive side-channel attacks, like power or electromagnetic (EM) analysis~\cite{cryptoKocherJJ99,esmartQuisquaterS01}, are among the most serious threats for implementations of cryptographic algorithms.
A common algorithmic countermeasure strategy against these attacks is the combination of masking against power analysis with redundant computation against fault attacks.
\emph{Masking} is a secret-sharing technique where one splits a cryptographic computation into $d+1$ random shares.
This technique ensures that the observation of up to $d$ intermediate values of that masked computation
does not reveal any information about native (unmasked) values~\cite{cryptoIshaiSW03,tchesGrossIB18,chesGrossM17,eurocryptBartheDFGSS17,DBLP:journals/tches/FaustGPPS18}.
\emph{Redundant computation} tries to prevent the release of faulty cryptographic computations caused by environmental influences or malicious tampering such as voltage glitches, lasers, or rapid temperature variations.
Without this countermeasure, an attacker with access to faulty computations can learn information about the used cryptographic key in many different ways~\cite{pieeeBar-ElCNTW06,DBLP:conf/cardis/HutterS13,fdtcFuhrJLT13}.

Researchers long believed that the combination of redundancy and masking could adequately deal with active and passive implementation attacks.
However, it was recently shown that when using \emph{statistical ineffective fault attacks} (SIFA), even such protected cryptographic implementations are vulnerable to rather straightforward implementation attacks~\cite{DBLP:journals/tches/DobraunigEKMMP18,asiacryptDobraunigEGMMP18,iacrDobraunigMMP18}. 
The key observation behind SIFA attacks is that a cryptographic key may correlate with the suppression of a faulted cryptographic computation. 
Thus, the attacker can obtain information about this key by observing whether the output of a faulted cryptographic computation is suppressed by a redundancy countermeasure or not.

For example, if a 1-bit signal carries a secret value, and the attacker can force this signal to zero, they can learn the secret value by observing whether or not this fault is detected.
While this simplified example is obvious, SIFA is interesting because it works even if the fault injection targets just one share of a masked secret.
In fact, SIFA is exploitable even if the attacker does not know the exact effect of a fault injection on the faulted value~\cite{asiacryptDobraunigEGMMP18}.

Most proposed mitigation techniques against SIFA so far use error correction, which is however costly when combined with masking~\cite{DBLP:conf/ctrsa/DhoogheN20,DBLP:journals/iacr/SahaJRCBM19}.
Another recently proposed SIFA mitigation tries to solve this issue with a careful combination of redundancy, masking, and reversible computing~\cite{DBLP:journals/tches/DaemenDEGMP20}, achieving protection against SIFA without significant overheads.
The authors give detailed \emph{circuit} descriptions of protected cipher components that can be mapped into concrete software or hardware implementations. 
However, even minor modifications of the circuit description due to human error, compilers, or synthesis tools, although preserving functional equivalence, may make the circuit vulnerable to SIFA.
Consequently, there is a high demand for tooling that can support designers in building efficient
cryptographic implementations resistant against power analysis and fault attacks, including SIFA.

\subsection{Related Work}

The empirical and formal verification of power analysis and fault attack countermeasures is
an already well established topic in the cryptographic research community~\cite{DBLP:conf/icecsys/ArribasNR18,barthe2018maskverif,DBLP:conf/eurocrypt/BloemGIKMW18,Coco,DBLP:conf/tacas/GaoXZSC19,DBLP:journals/tosem/GaoZSW19,DBLP:conf/asiacrypt/KnichelS020}.
On a conceptual level, the verification of masking countermeasures --- ensuring that individual computations
are unrelated to any cryptographic secret --- does perform statistical independence checks
that could also be adapted for verifying SIFA protection, i.e., that cryptographic secrets do not correlate
with the suppression of a faulted cryptographic computation.
However, in the following we argue that such existing tools either cannot be easily adapted for SIFA verification, or would come with performance overheads that make them unattractive for practical use.

Tools like \rebecca/~\cite{DBLP:conf/eurocrypt/BloemGIKMW18} and its successor \coco/~\cite{Coco} use correlation tracking to show statistical independence in (sequential) masked hardware circuits. 
Although their method ignores the \emph{strength} and \emph{sign} of correlations for performance reasons, the remaining information is still sufficient to show standard probing resistance of masked circuits.
However, these approximations are not applicable for SIFA verification.
Since \rebecca/ and \coco/ do not track the \emph{sign} of correlations, there is no way to distinguish the correlation sets of a negated value from a non-negated value.
Due to the nature of bit-flip faults, this method leads to falsely reported leaks due to the structure of the fault-detection mechanism.
Similarly, tools like \texttt{maskVerif}~\cite{barthe2018maskverif} rely on security proofs for a gate's input signals to prove the gate's security. 
According to our investigation, since the fault-detection mechanism combines the shares in its sub-formulas, a leakage report is triggered even though the value cannot be observed.

Exact methods like SILVER~\cite{DBLP:conf/asiacrypt/KnichelS020} use some form of model counting to track exact probability distributions of values within masked circuits and check whether the correlation strength is zero for all secret values. 
These methods could be adopted for SIFA verification, e.g., by using a strategy as outlined in Figure~\ref{fig:redundant-circ} but will lead to verification runtimes significantly higher compared to the approach that we will present in this paper.

Besides masking verification tools, there also exists VerFI~\cite{DBLP:journals/iacr/ArribasWMN19}, a verification tool dedicated to fault attacks that, amongst others, does have the capability to verify SIFA protection of a given circuit in certain scenarios.
More precisely, VerFI can detect SIFA vulnerability of a given circuits using an empirical and simulation-based approach that essentially checks if either (1) all fault injections are being corrected through error correction methods, or (2) all fault injections are being detected via redundancy methods.
This empirical approach can be used for error-correction-based SIFA countermeasures, however, VerFI is not suited for the verification of, e.g., the more efficient SIFA countermeasure design by Daemen~\etal/\cite{DBLP:journals/tches/DaemenDEGMP20} that does not need be able to correct any possible fault injection.

\subsection{Contribution}

The contribution of this paper is threefold and consists of a method and its implementation, its evaluation, and resulting SIFA-resistant circuit artifacts.

\para{Method.} We present a formal verification approach to determine whether a masked redundant cipher implementation is SIFA resistant within a well-defined attacker model.
Our verification approach checks whether the output of the fault-detection mechanism correlates with secrets used in the computation.
We present three properties and their respective checking methods that serve as sufficient conditions for SIFA protection.
\emph{(Incompleteness):} If a function $\delta$ does not functionally depend on all shares of a secret $s$, it cannot leak the secret.
\emph{(Hiding):} If a function $\delta$ can be written as $m \oplus  \delta'$, where $m$ is a uniformly distributed random variable and $\delta'$ is functionally independent of $m$, $\delta$ does not leak information about any secrets.
\emph{(Inferred independence):} For a function $\delta = \bigvee_i \delta_i$, if all linear combinations of its partial functions $\delta_i$ are statistically independent of a secret $s$, $\delta$ cannot leak the secret $s$.
We present an algorithm that uses these sufficient but not necessary conditions to prove the security of circuits.
Our tool \tool/ implements this algorithm and is, to our knowledge, the first tool for formal verification of SIFA resistance of masked redundant circuits.

\para{Evaluation.} We provide an experimental evaluation of our method. 
Because the sufficient conditions may not be able to prove SIFA resistance, we show in our experimental section that the approach gives precise results for a representative range of secure circuits.
If \tool/ cannot prove resistance, it provides fault locations that might leak information about the secrets. We show that \tool/ can accurately prove security or find bugs in S-Boxes, the non-linear parts of cryptographic implementations, in minutes or even seconds. With respect to SIFA verification, masked linear layers do not need any further analysis as fault injections in these components are not exploitable with SIFA.
Ultimately, we give practical examples illustrating that, even when a design is secure against SIFA on paper, vulnerabilities may arise as a result of simple compiler/synthesis optimizations, which can then however be identified with \tool/.

\para{Artifacts.} As a direct result of this work, we present the first SIFA-resistant Verilog implementations of Daemen~\etal/\cite{DBLP:journals/tches/DaemenDEGMP20} designs for a masked AES S-Box, the \keccak/ $\chi_3$ S-Box, and all classes of quadratic 4-bit S-Boxes.

\section{Preliminaries} \label{sec:prelim}

\para{Masking} is an algorithmic countermeasure that, while primarily intended to prevent power analysis attacks, also plays an essential role in SIFA attacks.
In a masked cipher implementation, each input, output, and intermediate variable is split into $d+1$ \emph{shares} so that their \textsc{Xor} is equal to the original \emph{native variable}~\cite{cryptoIshaiSW03}. 
In Boolean masking, a native variable $s$ is split random shares $s_0 \hdots s_d$ that satisfy $s = s_0 \oplus \hdots \oplus s_{d}$. 
As long as an attacker cannot observe a set of values statistically dependent on all $d+1$ shares of a native value, the computation is secure against classical power analysis techniques.
Dealing with linear functions is trivial as they can be computed on each share individually. 
However, implementing masking for non-linear functions (S-Boxes) requires computations on all shares, which is more challenging to implement securely and correctly, and thus the main interest in the literature.

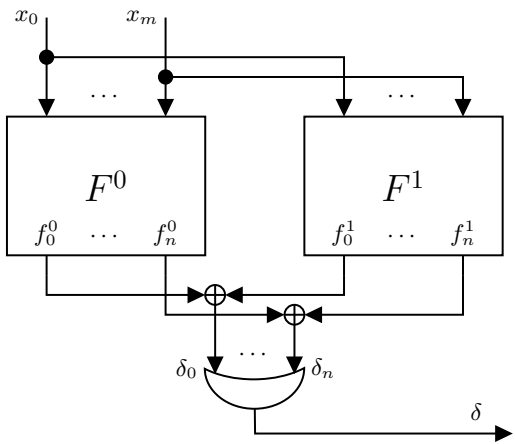
\begin{figure}[t]
\centering
\tikzset{every picture/.style={line width=0.75pt}} 

\begin{tikzpicture}[x=0.75pt,y=0.75pt,yscale=-1,xscale=1]

\draw    (130,40) -- (130,87) ;
\draw [shift={(130,90)}, rotate = 270] [fill={rgb, 255:red, 0; green, 0; blue, 0 }  ][line width=0.08]  [draw opacity=0] (8.93,-4.29) -- (0,0) -- (8.93,4.29) -- cycle    ;
\draw    (190,40) -- (190,87) ;
\draw [shift={(190,90)}, rotate = 270] [fill={rgb, 255:red, 0; green, 0; blue, 0 }  ][line width=0.08]  [draw opacity=0] (8.93,-4.29) -- (0,0) -- (8.93,4.29) -- cycle    ;
\draw    (130,60) -- (280,60) -- (280,87) ;
\draw [shift={(280,90)}, rotate = 270] [fill={rgb, 255:red, 0; green, 0; blue, 0 }  ][line width=0.08]  [draw opacity=0] (8.93,-4.29) -- (0,0) -- (8.93,4.29) -- cycle    ;
\draw [shift={(130,60)}, rotate = 0] [color={rgb, 255:red, 0; green, 0; blue, 0 }  ][fill={rgb, 255:red, 0; green, 0; blue, 0 }  ][line width=0.75]      (0, 0) circle [x radius= 3.35, y radius= 3.35]   ;
\draw    (190,70) -- (340,70) -- (340,87) ;
\draw [shift={(340,90)}, rotate = 270] [fill={rgb, 255:red, 0; green, 0; blue, 0 }  ][line width=0.08]  [draw opacity=0] (8.93,-4.29) -- (0,0) -- (8.93,4.29) -- cycle    ;
\draw [shift={(190,70)}, rotate = 0] [color={rgb, 255:red, 0; green, 0; blue, 0 }  ][fill={rgb, 255:red, 0; green, 0; blue, 0 }  ][line width=0.75]      (0, 0) circle [x radius= 3.35, y radius= 3.35]   ;
\draw    (130,170) -- (130,180) -- (207,180) ;
\draw [shift={(210,180)}, rotate = 180] [fill={rgb, 255:red, 0; green, 0; blue, 0 }  ][line width=0.08]  [draw opacity=0] (8.93,-4.29) -- (0,0) -- (8.93,4.29) -- cycle    ;
\draw    (280,170) -- (280,180) -- (223,180) ;
\draw [shift={(220,180)}, rotate = 360] [fill={rgb, 255:red, 0; green, 0; blue, 0 }  ][line width=0.08]  [draw opacity=0] (8.93,-4.29) -- (0,0) -- (8.93,4.29) -- cycle    ;
\draw    (190,170) -- (190,190) -- (247,190) ;
\draw [shift={(250,190)}, rotate = 180] [fill={rgb, 255:red, 0; green, 0; blue, 0 }  ][line width=0.08]  [draw opacity=0] (8.93,-4.29) -- (0,0) -- (8.93,4.29) -- cycle    ;
\draw    (340,170) -- (340,190) -- (263,190) ;
\draw [shift={(260,190)}, rotate = 360] [fill={rgb, 255:red, 0; green, 0; blue, 0 }  ][line width=0.08]  [draw opacity=0] (8.93,-4.29) -- (0,0) -- (8.93,4.29) -- cycle    ;
\draw   (210,180) .. controls (210,177.24) and (212.24,175) .. (215,175) .. controls (217.76,175) and (220,177.24) .. (220,180) .. controls (220,182.76) and (217.76,185) .. (215,185) .. controls (212.24,185) and (210,182.76) .. (210,180) -- cycle ; \draw   (210,180) -- (220,180) ; \draw   (215,175) -- (215,185) ;
\draw    (234.99,237.46) -- (235.14,249.96) -- (362.14,249.96) ;
\draw [shift={(365.14,249.96)}, rotate = 180] [fill={rgb, 255:red, 0; green, 0; blue, 0 }  ][line width=0.08]  [draw opacity=0] (8.93,-4.29) -- (0,0) -- (8.93,4.29) -- cycle    ;
\draw    (215,185) -- (215,217) ;
\draw [shift={(215,220)}, rotate = 270] [fill={rgb, 255:red, 0; green, 0; blue, 0 }  ][line width=0.08]  [draw opacity=0] (8.93,-4.29) -- (0,0) -- (8.93,4.29) -- cycle    ;
\draw    (255,195) -- (255,217) ;
\draw [shift={(255,220)}, rotate = 270] [fill={rgb, 255:red, 0; green, 0; blue, 0 }  ][line width=0.08]  [draw opacity=0] (8.93,-4.29) -- (0,0) -- (8.93,4.29) -- cycle    ;
\draw   (110,90) -- (210,90) -- (210,160) -- (110,160) -- cycle ;
\draw    (130,160) -- (130,170) ;
\draw    (190,160) -- (190,170) ;

\draw   (250,190) .. controls (250,187.24) and (252.24,185) .. (255,185) .. controls (257.76,185) and (260,187.24) .. (260,190) .. controls (260,192.76) and (257.76,195) .. (255,195) .. controls (252.24,195) and (250,192.76) .. (250,190) -- cycle ; \draw   (250,190) -- (260,190) ; \draw   (255,185) -- (255,195) ;
\draw   (259.9,216.91) .. controls (259.99,228.16) and (248.83,237.36) .. (234.99,237.46) .. controls (221.15,237.56) and (209.86,228.53) .. (209.78,217.28) .. controls (216.36,220.65) and (225.19,222.68) .. (234.88,222.61) .. controls (244.58,222.54) and (253.38,220.38) .. (259.9,216.91) -- cycle ;
\draw   (260,90) -- (360,90) -- (360,160) -- (260,160) -- cycle ;
\draw    (280,160) -- (280,170) ;
\draw    (340,160) -- (340,170) ;

\draw (235,210) node    {$\cdots $};
\draw (128,40) node [anchor=east] [inner sep=0.75pt]    {$x_{0}$};
\draw (188,40) node [anchor=east] [inner sep=0.75pt]    {$x_{m}$};
\draw (207.78,217.28) node [anchor=east] [inner sep=0.75pt]  [font=\small]  {$\delta _{0}$};
\draw (261.9,216.91) node [anchor=west] [inner sep=0.75pt]  [font=\small]  {$\delta _{n}$};
\draw (346.78,238.82) node  [font=\small]  {$\delta $};
\draw (130,156.6) node [anchor=south] [inner sep=0.75pt]  [font=\small]  {$f^{0}_{0}$};
\draw (190,156.6) node [anchor=south] [inner sep=0.75pt]  [font=\small]  {$f^{0}_{n}$};
\draw (160,150.5) node    {$\cdots $};
\draw (160,125) node  [font=\Large]  {$F^{0}$};
\draw (160,80) node    {$\cdots $};
\draw (310,80) node    {$\cdots $};
\draw (310,125) node  [font=\Large]  {$F^{1}$};
\draw (310,150.5) node    {$\cdots $};
\draw (340,156.6) node [anchor=south] [inner sep=0.75pt]  [font=\small]  {$f^{1}_{n}$};
\draw (280,156.6) node [anchor=south] [inner sep=0.75pt]  [font=\small]  {$f^{1}_{0}$};

\end{tikzpicture}
\caption{A redundant computation with inputs $x_0, \ldots, x_m$, which are passed to both computation instances $F^0$ and $F^1$. 
The disjunction of differences $\delta_0, \ldots, \delta_n$ is used to determine whether there was a fault in one of the computation instances.}
\label{fig:redundant-circ}
\vspace{-0.5cm}
\end{figure}

\para{Redundant computation} is an implementation-level fault attack countermeasure for cryptographic computations. 
The main idea is to perform the same computation several times and release a result only if the redundant computations match. 
This check prevents cases where an attacker forces faults in the computation, leading to incorrect results that correlate with native secrets~\cite{pieeeBar-ElCNTW06}. 
Figure~\ref{fig:redundant-circ} shows the structure of a fault detection mechanism for redundant computations.
If an attacker introduces a fault and the outputs do not match, output $\delta$ signals the faults and prevents the release of the result.

\para{Statistical ineffective fault attacks (SIFA),} first presented at CHES 2018 by Dobraunig~\etal/, is a relatively new type of fault attack technique capable of circumventing common fault/power analysis countermeasures, while being applicable to a wide variety of block ciphers or AEAD schemes~\cite{DBLP:journals/tches/DobraunigEKMMP18,iacrDobraunigMMP18,DBLP:conf/host/RamezanpourAD19,asiacryptDobraunigEGMMP18}.
When performing SIFA, an attacker calls a cryptographic operation (e.g. block cipher) with varying inputs, injects a fault during each of the computations, and only collects outputs in cases where the fault injection did not cause a faulty computation result (i.e. the output is not suppressed).
This \emph{filtered} set of outputs can then be used to perform a key recovery attack on a block cipher as follows.

A typical block cipher design of an iterated round function, consisting of a linear and non-linear layer, that mixes the current state with the cryptographic key such that in the end, each bit of the block cipher output is uniformly distributed.
If we now consider, e.g., an AND computation that occurs in the non-linear layer of a (later) round function,
one can observe that a fault-induced difference in one operand only propagates to the AND output if
the other operand is '1'.
Hence, if an attacker repeatedly calls a block cipher with varying inputs, while injecting the same difference
in each computation, and only collecting outputs that are correct (not suppressed), a certain intermediate
value should show a bias towards '0'.
Given such a set of faulted but correct block cipher outputs, an attacker can now make a partial key guess
of the last round key and calculate back to the faulted operation for each collected output (ciphertext).
If the partial key guess was correct, the observed distribution of an intermediate value at that location should be biased. Otherwise, if the observed distribution is uniform, the key guess was wrong.
For a more complete attack description targeting the AES-128 block cipher we refer to the description in~\cite{asiacryptDobraunigEGMMP18}.

If we now additionally consider masked implementations where each intermediate value is split into multiple random shares, filtering outputs based on the operand of one AND gate is not sufficient anymore.
In fact, for SIFA to work in masked scenarios, the attacker needs to work with fault inductions that
cause a difference that propagates into multiple AND gates that use the shares of one native value
as other operands.
We show this with a small example inspired by Daemen~\etal/\cite{DBLP:journals/tches/DaemenDEGMP20}.

\begin{figure}[t]
\centering
\includegraphics[width=0.4\textwidth]{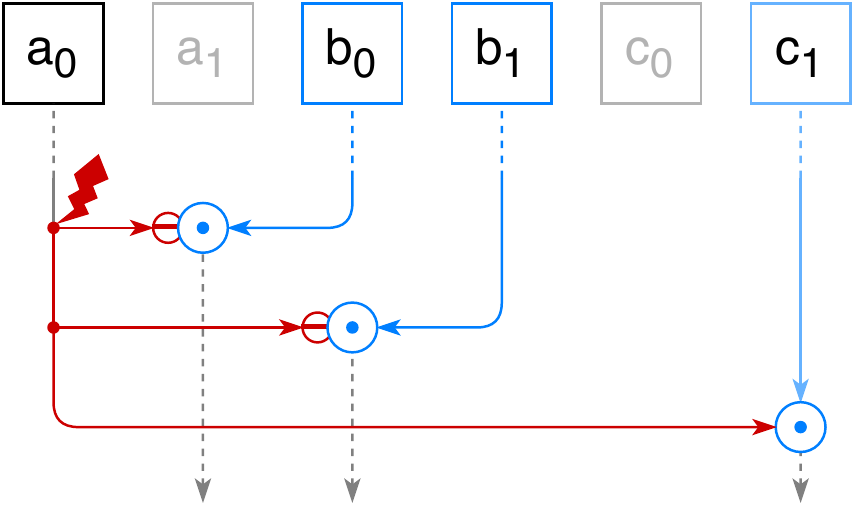}
\caption{Simplified example of SIFA against masked $\chi_3$ using two shares. The induced difference cancels out, and the attacker learns $b_0 \lor b_1 \lor c_1$. 
\ifdefined\print
\else
A complete depiction is given in Appendix~\ref{apx:sifa_big_2}.
\fi
}
\label{fig:sifa_small}
\end{figure}

\begin{example}
Consider a masked S-Box implementation that operates on shared inputs and outputs.
For simplicity, assume that we repeatedly call this S-Box with uniformly distributed inputs and observe the corresponding outputs. 
Since an S-Box is a bijective function, uniformly distributed inputs should give uniformly distributed outputs. 
Figure~\ref{fig:sifa_small} shows a reduced depiction of a masked $\chi_3$ S-Box, the smaller version of $\chi_5$, which is used in \keccak/ (SHA-3).
The S-Box takes a 3-bit input, represented by bits $a$, $b$ and $c$. Therefore a first-order masked version of $\chi_3$ takes the bits $a_0, a_1, b_0, b_1, c_0, c_1$ as input, with $a = a_0 \oplus a_1$, etc..
If we  assume a fault  targeting $a_0$ at the specified location,  the induced bit-difference propagates into three \textsc{And}-gates that take the bits $b_0$, $b_1$, and $c_1$ as the other inputs. In this case, the bit-difference cancels out and produces a value $\delta = b_0 \lor b_1 \lor c_1$. When a fault is not detected, an attacker knows that $b_0$, $b_1$, and $c_1$ are all zero, and therefore, $b$ is zero as well. 
In this concrete case, the attacker uses a fault injection to filter out computations where the distribution of $b$ is biased, and uses them to recover the key
\end{example}

\para{Efficient SIFA countermeasures}
were presented at CHES~2020~\cite{DBLP:journals/tches/DaemenDEGMP20}. Their SIFA mitigation strategy has almost no overhead and builds upon a careful combination of masking, redundant computation, and reversible computing.
They show that, by building non-linear operations from incomplete and invertible building blocks, they achieve implementations where a single fault in the computation is either (1) not exploitable by SIFA, or (2) detectable via redundant computations.
This approach is comparably easy to implement for small S-Boxes and can also be extended to larger S-Boxes such as the AES S-Box.
\ifdefined\print
\else
An example of a masked $\chi_3$ implementation that is built from incomplete and invertible building blocks is given in Appendix~\ref{apx:sifa_big_3}.
\fi

\para{Boolean formulas}
are a symbolic composition of Boolean variables using logic operators.
For a propositional boolean formula $f$, we write $\fvar{f}$ to refer to the variables that occur in $f$. 
When clear from context, we write $V$ to denote a superset of all used variables, \ie/ $\fvar{f} \subseteq V$.
The \emph{partial evaluation} of $f$, where a variable $q$ is assigned a value $p \in \bools/$ is written as $f\sub{p}{q}$. 
Given a set of variables $Q$ and an assignment $\alpha : Q \rightarrow \bools/$, we write $f\app{\alpha}$ to denote the partial evaluation of $f$ where each variable in $Q$ is assigned according to $\alpha$. 

We say that a formula $f$ is \emph{functionally dependent} on a variable $x$ if and only if the concrete value of $x \in \bools/$ has an influence on the value of $f \in \bools/$. 
Henceforth, for a given formula $f$, we write $\ess{f} \subseteq \fvar{f}$ to denote the set of variables that $f$ functionally depends on.
That is, $x \in \ess{f}$ if and only if there exists $\alpha : \fvar{f} \setminus \left\{x\right\} \rightarrow \bools/$, such that $f[\alpha]\sub{\0/}{x} \oplus f[\alpha]\sub{\1/}{x} = \1/$.
The above property can be checked by a SAT solver.

To discuss information leakage caused by a fault, we first define what it means for a formula $f$ to contain information about another formula $g$.
We define the weight of a Boolean function as $\numass{V}{f} = \left|\left\{\alpha: V \rightarrow \bools/ \mid f[\alpha] = \1/\right\}\right|$.
Formulas $f$ and $g$ are \emph{statistically dependent} if and only if $\numass{V}{f \land g} \cdot \numass{V}{\lnot f} \neq \numass{V}{\lnot f \land g} \cdot \numass{V}{f}$. 
That is, regardless of the observed value of $f$, the proportion of assignments $\alpha$ for which $g\app{\alpha} = \1/$ is constant. 
\vspace{-0.2em}
\begin{example} \label{ex:sdep}
Let $V = \{a, b, c\}$ be a set of variables. 
Let $f = a \land b$, $g = \lnot a \lor c$, and $h = b \oplus c$ be Boolean formulas.
Formulas $f$ and $g$ are statistically dependent because $\numass{V}{f \land g}\cdot \numass{V}{\lnot f} = 6$ and $\numass{V}{\lnot f \land g} \cdot \numass{V}{f} = 10$. 
Indeed, if $f\app{\alpha} = \1/$, then probably $g\app{\alpha} = \0/$, whereas if $f\app{\alpha} = \0/$, then $g\app{\alpha} = \1/$ is just as likely as $g\app{\alpha} = \0/$. 
The formulas $f$ and $h$ are statistically independent because $\numass{V}{f \land h}\cdot \numass{V}{\lnot f} = 6$ and $\numass{V}{\lnot f \land h} \cdot \numass{V}{f} = 6$.
\end{example}

We say that a Boolean formula $f$ is \emph{balanced} if and only if $\numass{V}{f} = \numass{V}{\lnot f} = 2^{|V| - 1}$. 
A Boolean variable $x$, interpreted as a formula, is inherently balanced for any variable set $x \in V$ as there are $2^{|V|-1}$ assignments $\alpha : V \rightarrow \bools/$ with $\alpha(x) = \1/$.  
Lemma~\ref{lem:balance} states that this can be extended to functions of the form $f = x \oplus g$.

\begin{lemmarep} \label{lem:balance}
Let $f = x \oplus g$ be a Boolean formula with $x \notin \fvar{g}$.
We have that $f$ is balanced.
\end{lemmarep}
\begin{proof}
Formula $f$ is balanced if $\numass{V}{x \oplus g} = \numass{V}{x \land \lnot g} + \numass{V}{\lnot x \land g} = 2^{|V| - 1}$.
Since $x \notin \fvar{g}$, we know that $\numass{V}{x \land \lnot g} = \numass{\{x\}}{x} \cdot \numass{V \setminus \{x\}}{\lnot g}$ and similarly $\numass{V}{\lnot x \land g} = \numass{\{x\}}{\lnot x} \cdot \numass{V \setminus \{x\}}{g}$. 
Therefore, since $\numass{\{x\}}{x} = \numass{\{x\}}{\lnot x} = 1$, we have that
\begin{equation*}
\numass{V}{x \oplus g} = \numass{V \setminus \{x\}}{\lnot g} + \numass{V \setminus \{x\}}{g} = \numass{V \setminus \{x\}}{\1/} = 2^{|V \setminus \{x\}|} = 2^{|V| - 1}. \ \ \ \ \qed
\end{equation*}
\end{proof}

We measure the \emph{Boolean distance} of two formulas $f$ and $g$ as the number of assignments where their values are different. 
This is equivalent to the weight of their difference $\numass{V}{f \oplus g}$.
Lemma~\ref{lem:corr} states the connection between statistical independence and Boolean distance.

\begin{lemmarep} \label{lem:corr}
Let $f$ and $g$ be Boolean formulas and let $f$ be balanced.
Formulas $f$ and $g$ are statistically independent if and only if their difference is balanced.
\end{lemmarep}
\begin{proof}
We start from the definition of statistical independence between $f$ and $g$.
Since $f$ is balanced, we can simplify the condition.
\begin{align*}
\numass{V}{f \land g}~\numass{V}{\lnot f} &= \numass{V}{\lnot f \land g}~\numass{V}{f} \\
\numass{V}{f \land g} &= \numass{V}{\lnot f \land g}.
\end{align*}
We know that the assignments $\alpha$ that satisfy $f \land g$ are all those that satisfy $f$, without those that satisfy $f \land \lnot g$.
We use this to show that the difference of $f$ and $g$ must be balanced.
\begin{align*}
\numass{V}{f} - \numass{V}{f\land \lnot g} &= \numass{V}{\lnot f \land g} \\
\numass{V}{f\land \lnot g} + \numass{V}{\lnot f \land g} &= \numass{V}{f} \\
\numass{V}{f \oplus g} &= 2^{|V| - 1}. \ \ \ \ \ \qed
\end{align*}
\end{proof}

\section{Verification Method} \label{sec:verif}

In this section, we introduce a method for verifying resistance against SIFA.
That is, we show how to verify whether the fault-detection mechanism could give away information about native secrets processed by a software computation or hardware circuit.
Our method focuses on proving the statistical independence of the fault-detection value $\delta$ and any of the secrets $s \in \secrets/$.
We do not show this directly and instead try to prove the statistical independence using the \emph{incompleteness}, \emph{hiding}, and \emph{inferred statistical independence} properties we introduce in this section.
However, we first define the exact attack model considered in this verification approach.

\subsection{Attack Model} \label{sec:attack-model}

Formally proving resistance against SIFA requires a definition of the attacker's capabilities and the exact information they observe. 
We use an attack model that is very similar to the one introduced by Daemen~\etal/\cite{DBLP:journals/tches/DaemenDEGMP20}.
We consider redundant masked implementations of S-Boxes that the attacker can query.  
Figure~\ref{fig:redundant-circ} shows a diagram of such an implementation, where the outputs of the two computation units are used to compute the fault-detection value $\delta$.
With SIFA, the value of $\delta$ is the only information the attacker receives from the computation.
The goal of an attacker is to learn information about the native secret values processed by the computation.
The inputs of the computation are categorized as masks and secret shares.
In the rest of the section, we say that $\masks/$ is the set of mask variables, and $\secrets/$ is the set of formulas representing the secrets.
We, therefore, have the set of input variables $V = \masks/ \cup \bigcup_{s \in \secrets/} \fvar{s}$.

As SIFA is a fault attack, the attacker has the technical capabilities to introduce the fault that changes the value of an intermediate computation.
If we represent $\delta$ as a computational circuit, a fault modifies the output of precisely one logic gate used during the computation.
In our attack model, we consider faults that can negate the value of the gate by causing a bit-flip, which also captures many other fault models such as stuck-at faults for masked circuits~\cite{DBLP:journals/tches/DaemenDEGMP20}.
The attacker's goal is to find a fault location that would cause a statistical dependency between $\delta$ and one of the formulas $s \in \secrets/$.
Our verification does not currently take into account the possible effects of ``glitchy'' fault injections, i.e., faults with specific timing behavior that causes the output of gates to change (glitch) several times before reaching a stable logic state 
While it has been shown that such effects need to be taken into account for implementing masking correctly in hardware, it is currently not clear if, or to what extend, they are relevant for SIFA attacks in realistic attacker settings.

\begin{proposition} \label{prop:model} 
A computation with a fault-detection value $\delta$ is SIFA resistant against a fault-inducing attacker if $\delta$ is statistically independent of all native secrets $s \in S$.
\end{proposition}

\subsection{Incompleteness} \label{sec:incomplete}

First, we prove that a fault-detection formula $\delta$ that does not functionally depend on all shares of a secret $s$, cannot be statistically dependent on $s$. 
A syntactic version of this property is known as \emph{non-interference} in the literature~\cite{DBLP:conf/eurocrypt/BartheBDFGS15,barthe2018maskverif}.
Intuitively, if one of the shares is absent from the formula $\delta$, then an attacker cannot infer anything about $s$ without this missing piece of information.
Definition~\ref{def:incomplete} formally states this intuition of \emph{incomplete} secrets. 
Lemma~\ref{lem:incomplete} states that \emph{incompleteness} is sufficient for statistical independence.

\begin{definition} \label{def:incomplete}
Let $f$ be a formula, and $s$ be a secret represented by the formula $s_0 \oplus \ldots \oplus s_d$, where the shares $s_i$ are variables. 
We say that a secret $s$ is \emph{incomplete} in formula $f$ whenever $\ess{s} \not \subseteq \ess{f}$.
\end{definition}

\begin{lemmarep} \label{lem:incomplete}
Let secret $s = s_0 \oplus \ldots \oplus s_d$ be incomplete in the fault-detection formula $\delta$. 
Then $\delta$ and $s$ are statistically independent.
\end{lemmarep}
\begin{proof}
Since $s$ is balanced per Lemma~\ref{lem:balance},  $\delta$ and $s$ are statistically independent if their difference $\delta \oplus s$ is balanced per Lemma~\ref{lem:corr}.
Let $\delta' = \delta \app{\fvar{\delta} \setminus \ess{\delta} \mapsto \0/}$ be a simplified version of $\delta$ from which the functionally irrelevant variables have been removed.
Due to incompleteness, there must be an $s_i \in \ess{s}$ such that $s_i \notin \fvar{\delta' \oplus \bigoplus_{j \neq i} s_j}$, and therefore per Lemma~\ref{lem:balance}, $\delta \oplus s$ is balanced.
\qed
\end{proof}

\subsection{Hiding} \label{sec:hiding}

Assume that the formula $\delta$ is functionally dependent on all shares of a secret $s = s_0 \oplus \ldots \oplus s_d$, \ie/ $\ess{s} \subseteq \ess{\delta}$.
Incompleteness, as defined in Definition~\ref{def:incomplete}, is thus not fulfilled.
However, $\delta$ and $s$ could still be statistically independent.
Intuitively, if $\delta$ is balanced and masked by some uniformly random value, it cannot statistically correlate with any secret $s \in \secrets/$. 

\begin{definition} \label{def:hiding}
A uniformly random variable $x$ \emph{hides} a secret $s \in \secrets/$ in the error-detection formula $\delta$ whenever $\delta = x \oplus f$, with $x \notin \ess{s} \cup \ess{f}$.
\end{definition}

Not all variables can hide secrets. Masks hide secrets because they are uniformly random by definition.
Although individual shares $s_i$ of a secret $s \in \secrets/$ are guaranteed to be uniformly random, their corresponding native secrets are not. 
Consequently, when investigating the hiding property from Definition~\ref{def:hiding}, we only consider masks and shares of incomplete secrets in $\delta$, as stated in Lemma~\ref{lem:hiding}. 

\begin{lemmarep} \label{lem:hiding}
Let $\delta$ be a formula, $\secrets/'$ be the set of secrets that are incomplete in $\delta$, \ie/ $\secrets/'  = \{s \in \secrets/ \mid \ess{s} \cap \ess{\delta} \neq \emptyset\}$, $\masks/$ be the uniformly random mask variables, and $X$ be the union $X = \masks/ \cup \bigcup_{s \in \secrets/'} \ess{s}$.
If there exists an $x \in X$ that hides a secret $s \in \secrets/$, then $\delta$ and $s$ are statistically independent.
\end{lemmarep}
\begin{proof}
Because $x$ hides $s$ in $\delta$, we know that $\delta \oplus s = x \oplus f \oplus s$ is balanced according to Lemma~\ref{lem:balance}. As the difference of $\delta$ and $s$ is balanced, they are statistically independent according to Lemma~\ref{lem:corr}.
\qed
\end{proof}

Lemma~\ref{lem:fact} presents a method that tests whether the factorization needed for the \emph{hiding} property is possible. 
The method uses a SAT solver and is similar to the method that checks functional dependencies.

\begin{lemmarep} \label{lem:fact}
Let $f$ be a Boolean formula and $x \in \fvar{f}$ be a variable. 
Then $f = x \oplus f\sub{\0/}{x}$ if and only if $f\sub{\0/}{x} \oplus f\sub{\1/}{x} = \1/$.
\end{lemmarep}
\begin{proof}
We assume that $f = x \oplus f\sub{\0/}{x}$ and show $f\sub{\0/}{x} \oplus f\sub{\1/}{x} = \1/$:
\begin{equation*} 
f\sub{\0/}{x} \oplus f\sub{\1/}{x} = (x \oplus f\sub{\0/}{x})\sub{\0/}{x} \oplus (x \oplus f\sub{\0/}{x})\sub{\1/}{x} = \1/ .
\end{equation*}
Similarly, assume that $f\sub{\0/}{x} \oplus f\sub{\1/}{x} = \1/$. We prove $f = x \oplus f\sub{\0/}{x}$ by showing that the two sides are equal under all values of $x$:
\begin{equation*}
\begin{split}
(x \oplus f\sub{\0/}{x})\sub{\0/}{x} & = f\sub{\0/}{x}\\
(x \oplus f\sub{\0/}{x})\sub{\1/}{x} & = (x \oplus f\sub{\1/}{x} \oplus \1/)\sub{\1/}{x} = f\sub{\1/}{x} . \ \ \ \ \ \qed
\end{split}
\end{equation*}
\end{proof}

It is enough to find one uniformly random variable $x$ to show that $\delta$ is statistically independent of all secrets $s \in \secrets/$.
As discussed earlier, not all variables in $\fvar{\delta}$ are eligible for the hiding property.
Thus, our verification method only checks the hiding property after determining incomplete secrets first.

\subsection{Inferred Statistical Independence} \label{sec:infer}

Although incompleteness and hiding are enough in most cases, the structure of $\delta$ can make them inapplicable. 
Therefore, it is possible that $\delta$ functionally depends on some secret $s$, and no uniformly random value hides $s$ in $\delta$. Example~\ref{ex:delta-bad} illustrates this situation.
\begin{example} \label{ex:delta-bad}
Let $\delta$ be the fault-detection formula with $\delta = \delta_0 \lor \delta_1$, $\delta_0 = x \oplus s_0$ and $\delta_1 = y \oplus s_1$ be its sub-formulas, $\masks/ = \left\{x, y\right\}$ be the masks, and $s = s_0 \oplus s_1$ be a secret.
Formula $\delta$ is functionally dependent on both $s_0$ and $s_1$, since there are no assignments $\alpha : \fvar{\delta} \setminus \{s_i\} \rightarrow \bools/$ such that $\delta[\alpha]\sub{\0/}{s_i} \oplus \delta[\alpha]\sub{\1/}{s_i} = \1/$.
Similarly, $\delta$ cannot be factorized into either $\delta = x \oplus \delta\sub{\0/}{x}$ or $\delta = y \oplus \delta\sub{\0/}{y}$, so neither $x$ nor $y$ hide $s$. 
However $\delta$ is indeed statistically independent of $s$ because $\numass{\fvar{\delta}}{\delta \land s}\cdot\numass{\fvar{\delta}}{\lnot \delta} = \numass{\fvar{\delta}}{\lnot \delta \land s}\cdot\numass{\fvar{\delta}}{\delta} = 24$.
\end{example}

Therefore, because of the structure of the fault-detection formula $\delta$, there is a real possibility that the incompleteness and hiding checks are not sufficient to show that $\delta$ does not statistically depend on any secrets.
However, this can be mitigated by inferring whether $\delta$ is statistically independent of $s$ by looking at its sub-formulas $\delta_i$ instead.
Lemma~\ref{lem:or-corr} introduces a method for inferring the statistical independence of two Boolean formulas $f$ and $g$, where one has the topmost operation \textsc{Or}, just like $\delta$, and the other is a balanced function, just like a secret. This property is inspired by correlation propagation used in \rebecca/~\cite{DBLP:conf/eurocrypt/BloemGIKMW18}.

\newcommand{\vxa}[0]{\lnot a \land \lnot b \land \lnot g} 
\newcommand{\vxb}[0]{      a \land \lnot b \land \lnot g} 
\newcommand{\vxc}[0]{\lnot a \land       b \land \lnot g} 
\newcommand{\vxd}[0]{\lnot a \land \lnot b \land       g} 
\newcommand{\vxe}[0]{      a \land \lnot b \land       g} 
\newcommand{\vxf}[0]{      a \land       b \land \lnot g} 
\newcommand{\vxg}[0]{\lnot a \land       b \land       g} 
\newcommand{\vxh}[0]{      a \land       b \land       g} 

\begin{lemmarep} \label{lem:or-corr}
Let $f = a \lor b$ and $g$ be Boolean formulas with the variable sets $\fvar{f} \subseteq V$ and $\fvar{g} \subseteq V$. If $\0/$, $a$, $b$, and $a \oplus b$ are statistically independent of $g$, then $f$ is also statistically independent of $g$.
\end{lemmarep}
\begin{proof} 
First, $\0/$ and $g$ are statistically independent if and only if $g$ is balanced, so $\numass{V}{g} = \numass{V}{\lnot g} = 2^{|V| - 1}$. Since $g$ must be balanced, $a$, $b$, $a \oplus b$ are statistically independent of $g$ if and only if their differences are balanced per Lemma~\ref{lem:corr}, \ie/ 
$\numass{V}{a \oplus g} = \numass{V}{b \oplus g} = \numass{V}{a \oplus b \oplus g} = 2^{|V| - 1}$. Summing up all of these conditions and rearranging them yields:
\begin{align*}
\numass{V}{\lnot g} + \numass{V}{a \oplus g} + \numass{V}{b \oplus g} + \numass{V}{a \oplus b \oplus g} &= 2^{|V| + 1}\\
\numass{V}{\vxa{}} + 3\cdot\numass{V}{\vxb{}} + 3\cdot\numass{V}{\vxc{}}\  &+ \\
+ 3\cdot\numass{V}{\vxd{}} + \numass{V}{\vxe{}} + 3\cdot\numass{V}{\vxf{}}\  &+ \\
+ \numass{V}{\vxg{}} + \numass{V}{\vxh{}} &= 2 ^{|V| + 1} \\
3 \cdot \numass{V}{f \oplus g} + \numass{V}{\lnot(f \oplus g)} &= 2^{|V| + 1}\\
2 \cdot \numass{V}{f \oplus g} + 2^{|V|} &= 2 ^{|V| + 1} \\
\numass{V}{f \oplus g} &= 2^{|V| - 1}. \ \ \ \qed
\end{align*}
\end{proof}

Therefore, at least in the case where $\delta = \delta_0 \lor \delta_1$, we can infer that $\delta$ is statistically independent of a secret $s$, as long as $\delta_0$, $\delta_1$, and $s$ fulfill the conditions of Lemma~\ref{lem:or-corr}. Example~\ref{ex:delta-bad2} illustrates this.
\begin{example} \label{ex:delta-bad2}
Let $\delta$, $\delta_0$, $\delta_1$ and $s$ be as in Example~\ref{ex:delta-bad}. By Lemma~\ref{lem:balance}, $s$ is balanced. 
The hiding property applies for $\delta_0$, $\delta_1$ and $\delta_0 \oplus \delta_1$, where  $x$, $y$, and  $x \oplus y$ can be factorized out respectively. 
According to Lemma~\ref{lem:corr}, all of the prerequisites for Lemma~\ref{lem:or-corr} are met, so we are able to show that $\delta$ is indeed statistically independent of $s$, without testing the statistical independence definition explicitly.
\end{example}

However, in general, $\delta$ will be a formula of the form $\delta = \bigvee_{i=1}^{n} \delta_i$. 
Although it is possible to apply Lemma~\ref{lem:or-corr} recursively, it is not ideal because we run into the same problem we demonstrated in Example~\ref{ex:delta-bad}, just one recursive application later.
Luckily, Lemma~\ref{lem:or-corr} can be generalized to \textsc{Or} operations with multiple arguments, as shown in Theorem~\ref{thm:or-corr}.

\begin{theoremrep} \label{thm:or-corr}
Let $\Phi = \{\phi_1, \ldots, \phi_n\}$ be a set of Boolean formulas, $f = \bigvee_{i=1}^{n} \phi_i$ be their disjunction, $g$ be another Boolean formula, and $\fvar{f} \subseteq V$ and $\fvar{g} \subseteq V$ be their variables. 
If for all $\Psi \in \pow{\Phi}$, where $\pow{\cdot}$ is the power-set operation, $f' = \bigoplus_{\psi \in \Psi} \psi$ is statistically independent of $g$, then so is $f$.
\end{theoremrep}
\begin{proof}
The proof proceeds by induction over the size of $\Phi$, where we write $\Phi_n = \left\{\phi_1, \ldots, \phi_n\right\}$ when referring to the set with $n$ sub-formulas.
\emph{Basis.} 
The conditions are trivially fulfilled for $\Phi_1 = \{\phi_1\}$, whereas $\Phi_2 = \{\phi_1, \phi_2\}$ directly applies Lemma~\ref{lem:or-corr}.
\emph{Step.} We show that the conditions hold for $\Phi_{n+1}$, under assumption that they hold for $\Phi_n$. 
We want to apply Lemma~\ref{lem:or-corr} to $f = a \lor b$ where $a = \phi_{n+1}$ and $b = \bigvee_{i=1}^{n} \phi_i$. 
The statistical independence of $\0/$, $a$ and $b$ with $g$ is guaranteed by respectively $\emptyset \in \pow{\Phi_{n+1}}$, $\{\phi_{n+1}\} \in \pow{\Phi_{n+1}}$, and $\pow{\Phi_n} \subset \pow{\Phi_{n+1}}$ together with the induction hypothesis for $\Phi_n$.
Lastly, if $g$ is balanced and statistically independent of $\phi_{n+1}$ like demonstrated above, then $a \oplus b$ is statistically independent of $g$ if and only if $b$ is statistically independent of $a \oplus g$.
This is guaranteed by $\left\{\Psi \cup \{\phi_{n+1}\} \mid \Psi \in \pow{\Phi_n}\right\} \subset \pow{\Phi_{n+1}}$ and the induction hypothesis.
Therefore, if for all $\Psi \in \pow{\Phi_{n+1}}$ we have that $f' = \bigoplus_{\psi \in \Psi} \psi$ is statistically independent of $g$, then, according to Lemma~\ref{lem:or-corr} and the previous observations, $f = \bigvee_{i=1}^{n+1} \phi_i$ is also statistically independent of $g$.
\qed
\end{proof}

Theorem~\ref{thm:or-corr} suggests that if we prove that all linear combinations of the error lines $\delta_i$ are statistically independent of a secret $s$, then we have indirectly shown that their disjunction $\delta$ is also statistically independent of $s$.
Additionally, the condition of Theorem~\ref{thm:or-corr} can be further simplified because some of the linear combinations produced by $X \in \pow{\Phi}$ could be equivalent. Instead of considering $\Phi$, we could instead consider the maximal linearly independent subset of $\Phi$.

\begin{lemmarep} \label{lem:or-simp}
Let $\Phi$ and $g$ be as in Theorem~\ref{thm:or-corr}. Let $\Phi' \subseteq \Phi$ be a linearly independent subset of $\Phi$, \ie/ $\forall \phi \in \Phi'.\ \forall \Psi \subseteq \Phi' \setminus \{\phi\}.\ \phi \neq \bigoplus_{\psi \in \Psi} \psi$, and let $\Phi'$ be maximal, \ie/ $\forall \phi \in \Phi \setminus \Phi'.\ \exists \Psi \subseteq \Phi'.\ \phi = \bigoplus_{\psi \in \Psi} \psi$.
If for all $\Psi \subseteq \Phi'$, $\bigoplus_{\psi \in \Psi} \psi$ is statistically independent of $g$, then the same holds for all $\Psi \subseteq \Phi$.
\end{lemmarep}
\begin{proof}
The stated implication relies on the maximality of the linearly independent subset $\Phi'$. In terms of spaces, $\Phi'$ is a basis of $\Phi$, meaning that it can represent all elements of $\Phi$ as an \textsc{Xor} of a subset of $\Phi'$, \ie/ $\forall \phi \in \Phi.\ \exists \Psi \subseteq \Phi'.\ \phi = \bigoplus_{\psi \in \Psi} \psi$. For each $\Psi \subseteq \Phi$, we can construct $\Psi' \subseteq \Phi'$ such that $\bigoplus_{\psi \in \Psi} \psi= \bigoplus_{\psi' \in \Psi'} \psi'$.
\qed
\end{proof}

As stated in Lemma~\ref{lem:or-simp}, instead of considering all linear combinations in $\Phi$, it is sufficient to consider only linear combinations of its maximally linearly independent subset $\Phi'$ when applying Theorem~\ref{thm:or-corr}.
In many cases, this substantially reduces the number of checks our verification method performed.

\subsection{Approximating Statistical Independence}

Theorem~\ref{thm:or-corr}, together with the optimized condition from Lemma~\ref{lem:or-simp}, is powerful enough to show that, given the mentioned conditions for $\delta_i$, $\delta$ is statistically independent of a secret $s$.
The statistical independence of the linear combinations of $\delta_i$ can be shown using the \emph{incompleteness} and \emph{hiding} properties discussed in Sections~\ref{sec:incomplete}~and~\ref{sec:hiding}. 
However, issuing exponentially many satisfiability queries required by Theorem~\ref{thm:or-corr} is still undesirable.
Therefore, we introduce an over-approximation which only calls the SAT solver to perform factorization and functional dependency tests for each relevant $\delta_i$ with all variables in $\fvar{\delta_i}$.
We then use the gathered data to over-approximate the incompleteness and hiding properties for all linear combinations of $\delta_i$.

In general a Boolean formula $f$ can be rewritten as an equivalent formula $f = g \oplus h$.
Here $g = \bigoplus_{x \in X} x$ is the linear sub-formula where $X \subseteq \fvar{f}$ is a set of variable symbols for which Lemma~\ref{lem:fact} applies, \ie/ $f\sub{\0/}{x} \oplus f\sub{\1/}{x} = \1/$.
Consequently, $h$ is the remaining sub-formula of $f$, \ie/ $h = f\app{\alpha}$ where $\alpha : X \mapsto \0/$ assigns $\0/$ to all variables in $X$.
Henceforth, we write $\fact{f}$ to denote the maximal set of variables that can be factorized out of $f$ via Lemma~\ref{lem:fact}, \ie/ $\fact{f} = \left\{x \mid x \in \fvar{f}, f\sub{\0/}{x} \oplus f\sub{\1/}{x} = \1/\right\}$.
Furthermore, call $f = \flin{f} \oplus \fnl{f}$ the maximal factorization, where $\flin{f} = \bigoplus_{x \in \fact{f}} x$, $\fnl{f} = f\app{\alpha}$ and $\alpha : \fact{f} \mapsto \0/$.
Knowing both $\fact{f}$ and $\ess{f}$ allows us to perform easy hiding and incompleteness checks for $f$ against some linear formula $f'$.
Additionally, $\fact{\cdot}$ and $\ess{\cdot}$ allow us to approximate the maximal factorization for linear combinations $f = \bigoplus_{i = 1}^n \phi_i$, where $\phi_i$ themselves are also formulas.

\begin{lemmarep} \label{lem:cd-approx}
Let $f = \bigoplus_{i = 1}^n \phi_i$ be a formula with sub-formulas $\phi_i$. 
The variable set $\xfact{f} = {\bigtriangleup_{i=1}^n \fact{\phi_i}} \setminus \bigcup_{i=1}^n \ess{\fnl{\phi_i}}$ is an under-approximation of $\fact{f}$.
Similarly, the set $\xess{f} = {\bigtriangleup_{i=1}^n \fact{\phi_i}} \cup \bigcup_{i=1}^n \ess{\fnl{\phi_i}}$ is an over-approximation of $\ess{f}$.
\footnote{Operator $\triangle$ signifies symmetric difference: $A \triangle B = (A \cup B) \setminus (A \cap B)$}
\end{lemmarep}
\begin{proof}
We rearrange the linear and non-linear parts of $\phi_i$ into the formulas $g = \bigoplus_{i = 1}^{n} \flin{\phi_i}$ and $h = \bigoplus_{i = 1}^n \fnl{\phi_i}$, with $f = g \oplus h$. 
The function $g$ is linear by construction, and because of the properties of \textsc{Xor}, only variables that appear in an uneven number of sets $\fact{\flin{\phi_i}} = \fact{\phi_i}$ are are part of $\fact{g}$, and thus $\fact{g} = {\bigtriangleup_{i=1}^n \fact{\phi_i}}$. 
For the function $h$, we assume that no functional dependencies of any $\fnl{\phi_i}$ are lost, and thus get an over-approximation $\ess{h} \subseteq X = \bigcup_{i=1}^n \ess{\fnl{\phi_i}}$.
It follows that $\xfact{f} = \fact{g} \setminus X \subseteq \fact{f}$, as $\xfact{f}$ only contains variables that factorizable in $g$ and are not a functional dependency of $h$.
Along the same lines, we get $\ess{f} \subseteq \xess{f} = \fact{g} \cup X$ because it contains all variables except for those that are in not essential in either $g$ or $h$. \qed
\end{proof}

These two approximations are much easier to compute than the real variable sets $\fact{\delta}$ and $\ess{\delta}$.
Ideally, we first compute $\ess{\delta_i}$ and $\fact{\delta_i}$ for each of the fault-detection values $\delta_i$ using a SAT solver.
Afterward, when checking all their linear combinations, we only use fast set computation operations from Lemma~\ref{lem:cd-approx}.
Since $\xess{\cdot}$ is an over-approximation, it must contain all functional dependencies and possibly some spurious ones. 
If we show the incompleteness of a secret $s$ with $\xess{\cdot}$, we would have gotten the same result with $\ess{\cdot}$.
Similarly, $\xfact{\cdot}$ contains a subset of the variables that can be factorized out of the formula. 
It is still a factorization, although it is not guaranteed to be maximal like $\fact{\cdot}$.
Therefore, if we show that a secret is hidden by some uniformly random variable using $\xfact{\cdot}$, it is guaranteed to be hidden.

\subsection{Verification Algorithm}

In this section, we summarize how the verification algorithm works.
In particular, we focus on the order of checks performed by the algorithm and show how they correspond to the previous exposition.
As described in Section~\ref{sec:attack-model}, the attacker can introduce a fault in any sub-formula $\phi$ of $\delta$.
The verification method summarized in Algorithm~\ref{alg:fault-checker} is given the faulted $\delta$ and its sub-formulas $\delta_i$, the set of masks $\masks/$, and the set of formulas $\secrets/$ representing each secret as a linear combination of its shares.
The show algorithm considers only one fault at a time, and our tool \tool/ runs it separately for each possible fault location.

\begin{algorithm}[t!]
\SetKwInOut{Input}{Input}
\SetKwInOut{Output}{Output}
\SetKw{Secure}{secure}
\SetKw{Unknown}{unknown}
\SetKw{Continue}{continue}
\SetKw{Or}{or}
\Input{fault detection formulas $\{\delta_1, \ldots, \delta_n\}$, $\delta := \bigvee_{i=1}^n \delta_i$ \\
       masks $\masks/$, secrets $\secrets/ = \left\{s^1, \ldots, s^d\right\}$}
\Output{\Secure or \Unknown}
$R := \masks/$            \atcp{variables that hide}
$K := \emptyset$    \atcp{complete secrets}
\For{$s \in \secrets/$}
{
    \lIf(\btcp{mark as complete}){$\ess{s} \subseteq \ess{\delta}$}
    { $K := K \cup \{s\}$} 
    \lIf(\btcp{shares can hide}){$\ess{s} \not \subseteq \ess{\delta}$}
    { $R := R \cup \left(\ess{s} \cap \ess{\delta}\right)$}   
}
\lIf(\btcp{incomplete or hidden}){$K = \emptyset$ \Or $R \cap \fact{\delta} \neq \emptyset$}{\Return \Secure}
$G := \emptyset$ \atcp{basis of $\delta_i$ formulas}
\For{$i \in \{1,\ldots,n\}$}
{
\lIf(\btcp{include $\delta_i$ in basis $G$}){$\forall G' \subseteq G. \ \delta_i \neq \bigoplus_{g \in G'}\ g$}{$G := G \cup \{\delta_i\}$}
}
\For{$G' \subseteq G$}
{
    $\phi = \bigoplus_{g \in G'}\ g$ \atcp{comb. of sub-formulas}   
    \lIf(\btcp{no secrets complete}){$\forall s \in K. \ess{s} \not \subseteq \xess{\phi}$}{\Continue}
    \lIf(\btcp{secrets are hidden}){$R \cap \xfact{\phi} \neq \emptyset$}{\Continue}    
    \Return \Unknown \atcp{$\phi$ maybe dependent}
}
\Return \Secure \atcp{all $\phi$ independent}
\caption{\tool/ algorithm for verifying SIFA resistance}
\label{alg:fault-checker}
\end{algorithm}

First, the algorithm computes the set $K$ of complete secrets, \ie/ secrets for which $\delta$ functionally depends on all its shares.
Simultaneously, the algorithm computes the set $R$ of uniformly random values that contains all masks $\masks/$ and shares of incomplete secrets $s \notin K$.
In the rest of the algorithm, only values in $R$ can hide secrets.
If there are no complete secrets in $K$ or a uniformly random variable from $R$ can be factorized out of $\delta$ and hides all secrets in $K$, we know that $\delta$ is statistically independent of the secrets $\secrets/$.

Next, the algorithm computes a maximal linearly independent subset $G$ of fault-detection values $\delta_i$.
As discussed previously in Lemma~\ref{lem:or-corr}, it is sufficient to apply Theorem~\ref{thm:or-corr} to this subset when proving statistical independence. 
The algorithm computes the approximations $\xess{\phi}$ and $\xfact{\phi}$ for all possible linear combinations $\phi$ from $G$. 
It uses the approximations to check whether any of the secrets in $K$ are complete in $\xess{\phi}$, and if they are, whether any of the random values from $R$ appear in $\xfact{\phi}$ and hide them.
If we were able to show statistical independence of secrets for all $\phi$, Algorithm~\ref{alg:fault-checker} declares the computation secure for the given fault. 

\begin{theorem}
Algorithm~\ref{alg:fault-checker} is sound: if it returns \textbf{secure},  the analyzed fault in the attack model from Section~\ref{sec:attack-model} is not exploitable via SIFA. 
\qed
\end{theorem}

\section{Case Studies} \label{sec:eval}

This section evaluates our new verification approach against the secured implementations presented by Daemen~\etal/\cite{DBLP:journals/tches/DaemenDEGMP20}.
\tool/~\footnote{\tool/'s code is available at \url{https://extgit.iaik.tugraz.at/scos/danira}} uses the netlist of a combinatorial circuit as the input. It interprets the inputs as variables and the intermediate computations as Boolean formulas. 
From a theoretical standpoint, it does not matter whether the analyzed circuit has a state or not because we only consider the outputs after the computation finishes.

In the rest of this section, we consider the SIFA-resistant masked implementations of \keccak/ $\chi_3$, all classes of quadratic 4-bit S-Boxes, and an AES S-Box~\cite{DBLP:journals/tches/DaemenDEGMP20}. 
We argue that without a sophisticated verification method, it is extremely easy to introduce bugs that produce correct computations but break the theoretical SIFA-resistance guarantees. 

Finally, we summarize the performance of \tool/ on several versions of the same designs.

\subsection{Masked Keccak $\chi_3$} \label{sec:keccak}

The \keccak/ permutation $\chi_3$ is a simple circuit with three inputs and three outputs used in many lightweight ciphers.
Implementing a masked version is straightforward because of its low polynomial degree.
\ref{alg:keccak-chi} shows the masked computation of $\chi_3$ proposed by Daemen~\etal/\cite{DBLP:journals/tches/DaemenDEGMP20}.
The secrets processed by the circuit are $a = a_0 \oplus a_1$, $b = b_0 \oplus b_1$ and $c = c_0 \oplus c_1$, whereas $m_r$ and $m_t$ are used as uniformly random masks.
The results of the computation $r$, $s$, and $t$ are also split into two shares, respectively. 
The circuit was designed in such a way that the outputs are used for fault detection.
Given two redundant computations of \ref{alg:keccak-chi} with outputs $\{r_0, r_1, s_0, s_1, t_0, t_1\}$ and $\{r_0', r_1', s_0', s_1', t_0', t_1'\}$, the fault-detection values are defined as $\delta_1 = r_0 \oplus r_0', \ldots, \delta_6 = t_1 \oplus t_1'$.

\begin{procedure}[tb]
\SetKwInOut{Input}{Input}
\SetKwInOut{Output}{Output}
\Input{$\{a_0, a_1\}, \{b_0, b_1\}, \{c_0, c_1\}, \masks/ = \{m_r, m_t\}$}
\Output{$\{r_0, r_1\}, \{s_0, s_1\}, \{t_0, t_1\}$}
\addtolength{\linewidth}{-15pt}
\begin{multicols}{4}
$m_s := m_r \oplus m_t$\;
$x_0 := \lnot b_0 \land c_1$\; \label{alg:keccak-chi-bn00}
$x_2 := a_1 \land b_1$\;
$x_1 := \lnot b_0 \land c_0$\; \label{alg:keccak-chi-bn01}
$x_3 := a_1 \land b_0$\;
$r_0 := x_0 \oplus m_r$\;
$t_1 := x_2 \oplus m_t$\;
$r_0 := r_0 \oplus x_1$\;
$t_1 := t_1 \oplus x_3$\;
$x_0 := \lnot c_0 \land a_1$\;
$x_2 := b_1 \land c_1$\;
$x_1 := \lnot c_0 \land a_0$\;
$x_3 := b_1 \land c_0$\;
$s_0 := x_0 \oplus m_s$\;
$r_1 := x_2 \oplus m_r$\;
$s_0 := s_0 \oplus x_1$\;
$r_1 := r_1 \oplus x_3$\;
$x_0 := \lnot a_0 \land b_1$\;
$x_2 := c_1 \land a_1$\;
$x_1 := \lnot a_0 \land b_0$\;
$x_3 := c_1 \land a_0$\;
$t_0 := x_0 \oplus m_t$\;
$s_1 := x_2 \oplus m_s$\;
$t_0 := t_0 \oplus x_1$\;
$s_1 := s_1 \oplus x_3$\;
$r_0 := r_0 \oplus a_0$\;
$t_1 := t_1 \oplus c_1$\;
$s_0 := s_0 \oplus b_0$\;
$r_1 := r_1 \oplus a_1$\;
$t_0 := t_0 \oplus c_0$\;
$s_1 := s_1 \oplus b_1$\;
\end{multicols}
\caption{Chi3(): Implementation of a masked \keccak/ $\chi_3$ S-Box \cite{DBLP:journals/tches/DaemenDEGMP20}}
\label{alg:keccak-chi}
\end{procedure}

Each line of \ref{alg:keccak-chi} is a possible fault location according to our attack model in Section~\ref{sec:attack-model}.
Introducing a bit-flip fault means negating the result of one such line in one of the redundant computations.
Our verification method goes through each of the fault locations, negates the result at that point in the computation, and generates the fault-detection formulas $\delta_1, \ldots, \delta_6$.
We specify $\secrets/ = \{a, b, c\}$ and $\masks/ = \{m_r, m_t\}$, and run Algorithm~\ref{alg:fault-checker} to see if the considered fault could leak information about the secrets.

We implemented the netlist for \ref{alg:keccak-chi} manually, and \tool/ was able to verify that the design proposed in \cite{DBLP:journals/tches/DaemenDEGMP20} was indeed SIFA resistant.
However, when we synthesized an equivalent RTL design with Yosys, \tool/ reported that it could not prove SIFA resistance.
In the synthesized netlist, Yosys introduced a temporary gate $v_0 = \lnot b_0$ which it used to simplify Line~\ref{alg:keccak-chi-bn00} to $x_0 := v_0 \land c_1$ and Line~\ref{alg:keccak-chi-bn01} to $x_1 := v_0 \land c_0$.
Although this makes sense from an optimization perspective because it effectively reduces the size of the circuit by one gate, it breaks the SIFA resistance.
A fault at this new gate $v_0$ in the synthesized design is the same as two faults at Lines~\ref{alg:keccak-chi-bn00}~and~\ref{alg:keccak-chi-bn01}. 
As a result, $\delta$ becomes statistically dependent on $c$, which the attacker can exploit.
Unfortunately, this demonstrates that \emph{(1)} an analysis on the gate level is unavoidable and \emph{(2)} they must be implemented manually, as synthesis tools or compilers break SIFA resistance while maintaining functional correctness.

\subsection{Masked AES S-Box} \label{sec:aes}

Compared to $\chi_3$, the AES S-Box is a significantly more complex circuit of high polynomial degree.
The authors of the CHES~paper~\cite{DBLP:journals/tches/DaemenDEGMP20} propose a high-level sketch of a SIFA-resistant masked AES S-Box.
There are many ways to implement this high-level description and achieving SIFA resistance is not trivial.
After several failed attempts, we managed to implement a protected version of the proposed AES S-Box with the help of our new verification tool. 
We are convinced that correctly protecting a circuit as large as an AES S-Box is infeasible without the help of an automated verification method such as \tool/.

\subsection{Performance Evaluation}

This section gives a breakdown of \tool/'s performance on correctly (and incorrectly) protected implementations. 
We performed all experiments on a notebook with an eight-core \emph{Intel i7-8550U 1.8GHz} CPU and \emph{16} GiB of memory. 

\def\pcl/{perm.}

\begin{table}[t]
\center
\caption{Performance of \tool/ (D) and a modified version of SILVER~\cite{DBLP:conf/asiacrypt/KnichelS020} (S) for different  masked designs. 
Correct (incorrect) designs  are denoted by {\faCheck} ({\faRemove}).
In all cases, the reused gate was the exploitable  fault location.}
\begin{tabular*}{\textwidth}{l  @{\extracolsep{\fill}} rrrcrr}
\toprule
Design & Gates & $(\land)$ & $(\oplus)$ & Result & D (s) & S (s)\\
\midrule
\keccak/ $\chi_3$, full \ref{alg:keccak-chi} & 37 & 12 & 25 & \faCheck  & 0.06 & 0.24 \\
\keccak/ $\chi_3$, reuse $\lnot b_0$ & 36 & 12 & 24 & \faRemove & 0.05 & 0.07 \\
\keccak/ $\chi_3$, reuse $\lnot c_0$ & 36 & 12 & 24 & \faRemove & 0.06 & 0.12 \\
\keccak/ $\chi_3$, reuse $\lnot a_0$ & 36 & 12 & 24 & \faRemove & 0.06 & 0.18 \\

4-bit \pcl/ $\mathcal{Q}^4_{4}$~\cite{DBLP:journals/tches/DaemenDEGMP20}
& 10 & 4 & 6 & \faCheck & 0.03 & 0.10 \\
4-bit \pcl/ $\mathcal{Q}^4_{12}$~\cite{DBLP:journals/tches/DaemenDEGMP20}
& 20 & 8 & 12 & \faCheck & 0.05 & 0.16 \\
4-bit \pcl/ $\mathcal{Q}^4_{293}$~\cite{DBLP:journals/tches/DaemenDEGMP20}
& 30 & 12 & 18 & \faCheck & 0.05 & 0.23 \\
4-bit \pcl/ $\mathcal{Q}^4_{294}$~\cite{DBLP:journals/tches/DaemenDEGMP20}
& 30 & 12 & 18 & \faCheck & 0.04 & 0.21 \\
4-bit \pcl/ $\mathcal{Q}^4_{299}$~\cite{DBLP:journals/tches/DaemenDEGMP20}
& 50 & 20 & 30 & \faCheck & 0.07 & 0.41 \\
4-bit \pcl/ $\mathcal{Q}^4_{300}$~\cite{DBLP:journals/tches/DaemenDEGMP20}
& 36 & 12 & 24 & \faCheck & 0.06 & 0.26 \\

AES S-Box, reuse $g_{104}$
& 631 & 144 & 487 & \faRemove & 14.67 & 551.1 \\
AES S-Box, reuse $g_{240}$
& 631 & 144 & 487 & \faRemove & 83.28 & 1336.7 \\
AES S-Box, reuse $g_{360}$
& 631 & 144 & 487 & \faRemove & 135.04 & 1941.7 \\
AES S-Box, full~\cite{DBLP:journals/tches/DaemenDEGMP20} 
& 634 & 144 & 490 & \faCheck & 184.39 & 3297.4 \\
\bottomrule
\end{tabular*}\vspace{0.3em}
\label{tab:performance}
\vspace{-0.5cm}
\end{table}

As shown in Table~\ref{tab:performance}, \tool/ instantly verified (or falsified) all tested \keccak/ $\chi_3$ and quadratic 4-bit S-Box designs.
We also demonstrate that for \keccak/ $\chi_3$ and the AES S-Box, even one re-used gate leads to vulnerabilities.
\tool/ verifies the SIFA resistance of our implementation in about three minutes.
For the AES S-Boxes, \tool/ performs significantly better than a version SILVER~\cite{DBLP:conf/asiacrypt/KnichelS020} which we extended to verify SIFA resistance.
However, although this shows \tool/'s potential, our extension of SILVER with construct as shown in Figure~\ref{fig:redundant-circ} is not perfect and could be further improved by its authors.

In summary, the results of our experiments in Table~\ref{tab:performance} indicate that: (1) the over-approximation we introduce in this paper is strong enough to prove SIFA resistance for secure designs, and (2) our verification method applied by \tool/ is fast enough for complex masked implementations.

\section{Conclusion} \label{sec:conclusion}

Protecting masked implementations against SIFA is not straightforward.
Designers can make mistakes when implementing a specification that is supposed to be secure. 
Additionally, compilers and synthesis tools can introduce simplifications that break the SIFA-resistance guarantees.
\tool/ solves these problems using simple yet effective properties of redundant masked implementations to show whether they are SIFA resistant.
As demonstrated by our case studies, \tool/ is  able to verify designs that may be used in actual embedded systems. 
In cases where \tool/ cannot prove the security of a design, it gives a developer detailed debugging information about a problematic fault location.

\bibliography{bibliography}{}

\begin{thebibliography}{10}

\bibitem{DBLP:conf/icecsys/ArribasNR18}
V.~Arribas, S.~Nikova, and V.~Rijmen.
\newblock {VerMI}: Verification tool for masked implementations.
\newblock In {\em {ICECS}}, 2018.

\bibitem{DBLP:journals/iacr/ArribasWMN19}
V.~Arribas, F.~Wegener, A.~Moradi, and S.~Nikova.
\newblock Cryptographic fault diagnosis using {VerFI}.
\newblock {\em {IACR} Cryptol. ePrint Arch.}, 2019.

\bibitem{pieeeBar-ElCNTW06}
H.~Bar-El, H.~Choukri, D.~Naccache, M.~Tunstall, and C.~Whelan.
\newblock The sorcerer's apprentice guide to fault attacks.
\newblock {\em Proceedings of the {IEEE}}, 94(2), 2006.

\bibitem{barthe2018maskverif}
G.~Barthe, S.~Bela{\"{\i}}d, G.~Cassiers, P.~Alain Fouque, B.~Gr{\'{e}}goire,
  and F.-X. Standaert.
\newblock {maskVerif}: Automated verification of higher-order masking in
  presence of physical defaults.
\newblock In {\em ESORICS}, 2019.

\bibitem{DBLP:conf/eurocrypt/BartheBDFGS15}
G.~Barthe, S.~Bela{\"{\i}}d, F.~Dupressoir, P.{-}A. Fouque, B.~Gr{\'{e}}goire,
  and P.{-}Y. Strub.
\newblock Verified proofs of higher-order masking.
\newblock In {\em EUROCRYPT}, 2015.

\bibitem{eurocryptBartheDFGSS17}
G.~Barthe, F.~Dupressoir, S.~Faust, B.~Gr{\'{e}}goire, F.-X. Standaert, and
  P.-Y. Strub.
\newblock Parallel implementations of masking schemes and the bounded moment
  leakage model.
\newblock In {\em EUROCRYPT}, 2017.

\bibitem{cryptoBihamS97}
E.~Biham and A.~Shamir.
\newblock Differential fault analysis of secret key cryptosystems.
\newblock In {\em CRYPTO}, 1997.

\bibitem{DBLP:conf/eurocrypt/BloemGIKMW18}
R.~Bloem, H.~Gro{\ss}, R.~Iusupov, B.~K{\"{o}}nighofer, S.~Mangard, and
  J.~Winter.
\newblock Formal verification of masked hardware implementations in the
  presence of glitches.
\newblock In {\em EUROCRYPT}, 2018.

\bibitem{eurocryptBonehDL97}
D.~Boneh, R.~A. DeMillo, and R.~J. Lipton.
\newblock On the importance of checking cryptographic protocols for faults.
\newblock In {\em EUROCRYPT}, 1997.

\bibitem{DBLP:journals/tches/DaemenDEGMP20}
J.~Daemen, C.~Dobraunig, M.~Eichlseder, H.~Gro{\ss}, F.~Mendel, and R.~Primas.
\newblock Protecting against statistical ineffective fault attacks.
\newblock {\em TCHES}, 2020.

\bibitem{DBLP:conf/ctrsa/DhoogheN20}
S.~Dhooghe and S.~Nikova.
\newblock My gadget just cares for me -- how {NINA} can prove security against
  combined attacks.
\newblock In {\em {CT-RSA}}, 2020.

\bibitem{asiacryptDobraunigEGMMP18}
C.~Dobraunig, M.~Eichlseder, H.~Gro{\ss}, S.~Mangard, F.~Mendel, and R.~Primas.
\newblock {Statistical Ineffective Fault Attacks} on masked {AES} with fault
  countermeasures.
\newblock In {\em ASIACRYPT}, 2018.

\bibitem{DBLP:journals/tches/DobraunigEKMMP18}
C.~Dobraunig, M.~Eichlseder, T.~Korak, S.~Mangard, F.~Mendel, and R.~Primas.
\newblock {SIFA:} exploiting ineffective fault inductions on symmetric
  cryptography.
\newblock {\em TCHES}, 2018.

\bibitem{iacrDobraunigMMP18}
C.~Dobraunig, S.~Mangard, F.~Mendel, and R.~Primas.
\newblock Fault attacks on nonce-based authenticated encryption: Application to
  {Keyak} and {Ketje}.
\newblock In {\em SAC}, 2018.

\bibitem{DBLP:journals/tches/FaustGPPS18}
S.~Faust, V.~Grosso, S.~Merino~Del Pozo, C.~Paglialonga, and F.-X. Standaert.
\newblock Composable masking schemes in the presence of physical defaults {\&}
  the robust probing model.
\newblock {\em TCHES}, 2018.

\bibitem{fdtcFuhrJLT13}
T.~Fuhr, \'{E}. Jaulmes, V.~Lomn{\'{e}}, and A.~Thillard.
\newblock Fault attacks on {AES} with faulty ciphertexts only.
\newblock In {\em FDTC}, 2013.

\bibitem{DBLP:conf/tacas/GaoXZSC19}
P.~Gao, H.~Xie, J.~Zhang, F.~Song, and T.~Chen.
\newblock Quantitative verification of masked arithmetic programs against
  side-channel attacks.
\newblock In {\em TACAS}, 2019.

\bibitem{DBLP:journals/tosem/GaoZSW19}
P.~Gao, J.~Zhang, F.~Song, and C.~Wang.
\newblock Verifying and quantifying side-channel resistance of masked software
  implementations.
\newblock {\em TOSEM}, 28(3), 2019.

\bibitem{Coco}
B.~Gigerl, V.~Hadzic, R.~Primas, S.~Mangard, and R.~Bloem.
\newblock Coco: Co-design and co-verification of masked software
  implementations on {CPU}s.
\newblock In {\em {USENIX}}, 2021.

\bibitem{tchesGrossIB18}
H.~Gro{\ss}, R.~Iusupov, and R.~Bloem.
\newblock Generic low-latency masking in hardware.
\newblock {\em {IACR} Transactions on Cryptographic Hardware and Embedded
  Systems}, 2018.

\bibitem{chesGrossM17}
H.~Gro{\ss} and S.~Mangard.
\newblock Reconciling d+1 masking in hardware and software.
\newblock In {\em CHES}, 2017.

\bibitem{DBLP:conf/cardis/HutterS13}
M.~Hutter and J.-M. Schmidt.
\newblock The temperature side channel and heating fault attacks.
\newblock In {\em CARDIS}, 2013.

\bibitem{cryptoIshaiSW03}
Y.~Ishai, A.~Sahai, and D.~A. Wagner.
\newblock {Private Circuits}: Securing hardware against probing attacks.
\newblock In {\em CRYPTO}, 2003.

\bibitem{DBLP:conf/asiacrypt/KnichelS020}
D.~Knichel, P.~Sasdrich, and A.~Moradi.
\newblock {SILVER} - statistical independence and leakage verification.
\newblock In {\em ASIACRYPT}, 2020.

\bibitem{cryptoKocherJJ99}
P.~C. Kocher, J.~Jaffe, and B.~Jun.
\newblock Differential power analysis.
\newblock In {\em CRYPTO}, 1999.

\bibitem{esmartQuisquaterS01}
J.-J. Quisquater and D.~Samyde.
\newblock Electromagnetic analysis ({EMA}): Measures and counter-measures for
  smart cards.
\newblock In {\em E-smart}, 2001.

\bibitem{DBLP:conf/host/RamezanpourAD19}
K.~Ramezanpour, P.~Ampadu, and W.~Diehl.
\newblock A statistical fault analysis methodology for the {Ascon}
  authenticated cipher.
\newblock In {\em {HOST}}, 2019.

\bibitem{DBLP:journals/iacr/SahaJRCBM19}
S.~Saha, D.~Jap, D.~B. Roy, A.~Chakraborty, S.~Bhasin, and D.~Mukhopadhyay.
\newblock A framework to counter statistical ineffective fault analysis of
  block ciphers using domain transformation and error correction.
\newblock {\em {TIFS}}, 2020.

\end{thebibliography}
\bibliographystyle{plain}

\ifdefined\print
\else

\appendix

\newpage
\section{SIFA on masked $\chi_3$ S-Box}
\label{apx:sifa_big_2}
\begin{figure}[h]
\centering
\includegraphics[width=0.7\textwidth]{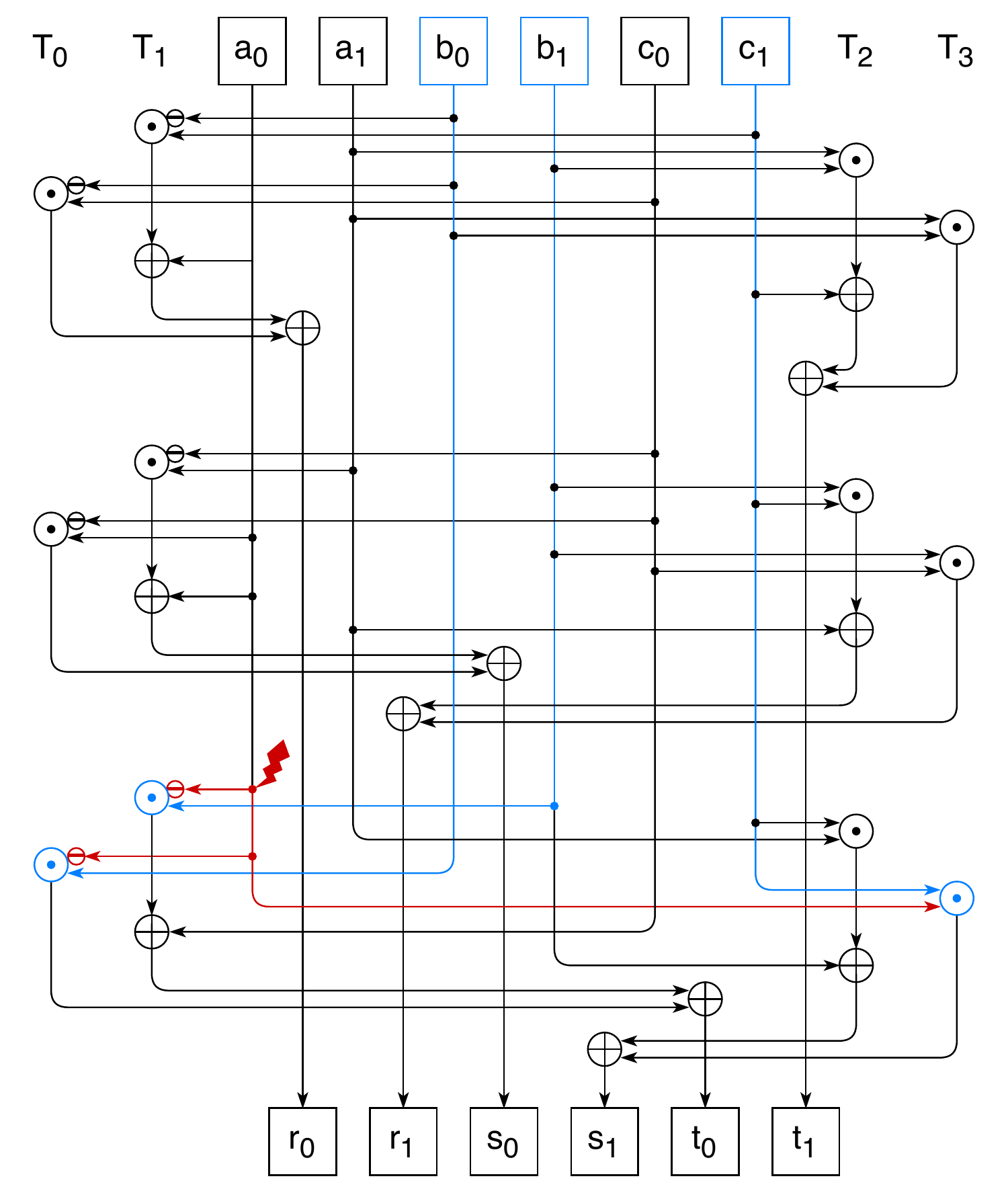}
\caption{SIFA against masked $\chi_3$ using 2 shares (from~\cite{DBLP:journals/tches/DaemenDEGMP20}). The induced difference cancels out, e.g., if the conrete values of $b_0, b_1$ and $c_1$ are all zero. If such a fault injection is not detected via redundant computation, the native value $b$ (i.e. $b_0 \oplus b_1$) at the input was zero (biased).}
\label{fig:sifa_big_2}
\end{figure}

\newpage
\section{SIFA protected masked $\chi_3$ S-Box}
\label{apx:sifa_big_3}
\begin{figure}[h]
\centering
\includegraphics[width=0.7\textwidth]{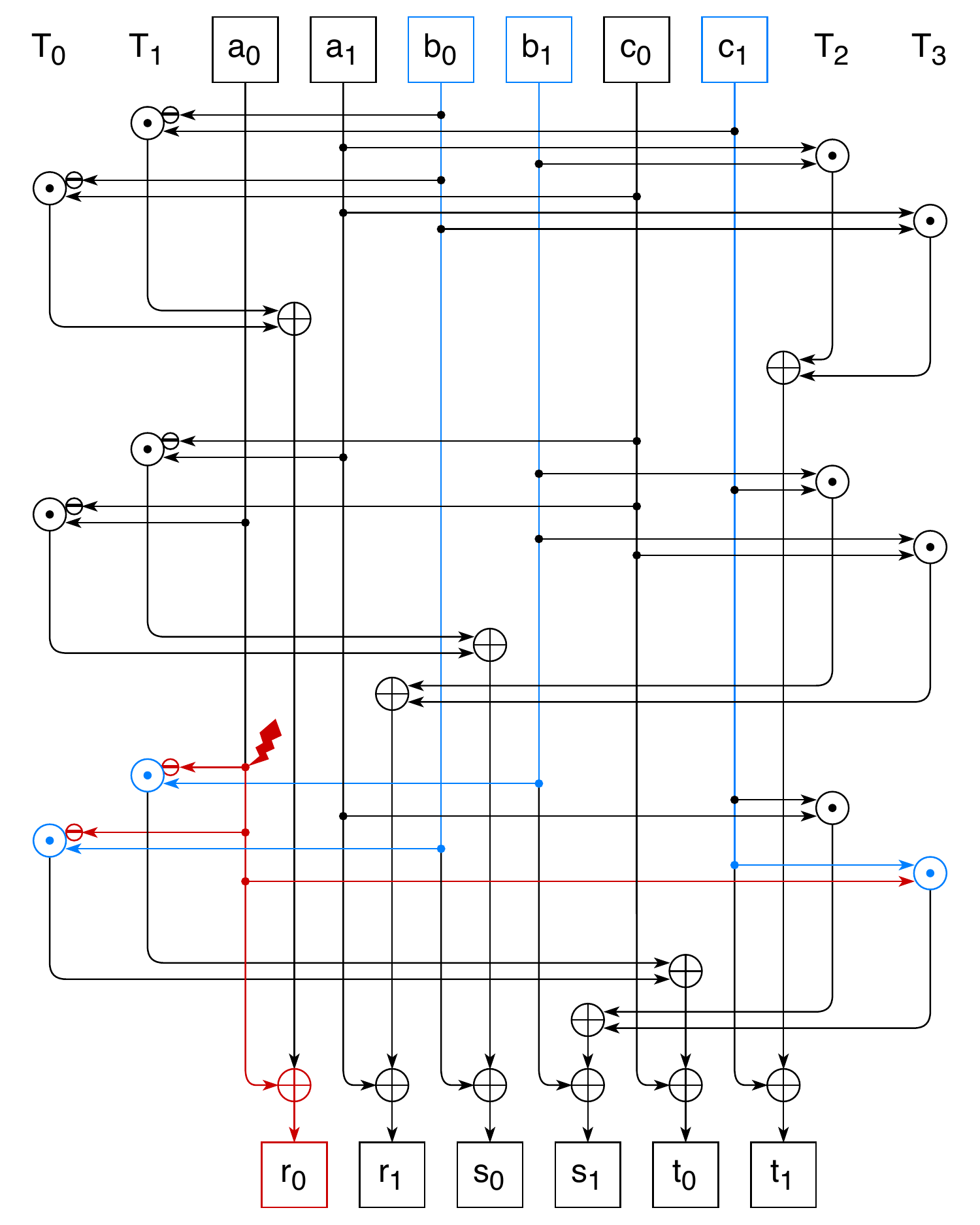}
\caption{SIFA protected masked $\chi_3$ using 2 shares (from~\cite{DBLP:journals/tches/DaemenDEGMP20}). For each possible fault location it either holds that (1) the induced difference is canceled by a set of signals that does not contain all shares of a native variable, or (2) the induced difference can be detected at the S-Box output via redundant computations. In this concrete example, the second property holds.}
\label{fig:sifa_big_3}
\end{figure}
\fi

\end{document}